\documentstyle[12pt]{article}
\begin{document}
%%%%%%%%%%%%%%%%%%%%%%%%%%%%%%%%%%%%%%%%%%%%%%%%%
\def\D{{\cal D}}
\def\R{{{\rm I} \! {\rm R}}}
\def\d{{{\rm \Delta} \!\!\!\! {\Delta}}}
\newcommand{\dt}{\displaystyle}
\newcommand{\btup}{{\bigtriangleup}}
\newcommand{\id}{{\rm id}}
\newcommand{\N}{{\bf\rm N}}
\newcommand{\I}{{\bf\rm I}}
\newcommand{\ctg}{{\rm ctg}}
\newcommand{\e}{{\rm e}}
\newcommand{\mod}{{\rm mod}}
\newcommand{\df}{\stackrel{df}{=}}
%%%%%%%%%%%%%%%%%%%%%%%%%%%%%%%%%%%%%%%%%%%%%%%%%%
%%%%%%%%%%%%%%%%%%%%%%%%%%%%%%%%%%%%%%%%%%%%%%%%%%
\vspace*{2cm}
\begin{center}
 \Huge\bf
Interpretation of intuitionistic solution of  the vacuum
Einstein  equations in smooth topos

\vspace*{0.25in}

\large

Alexander K.\ Guts, Artem A. Zvyagintsev
\vspace*{0.15in}

\normalsize

Department of Mathematics, Omsk State University \\
644077 Omsk-77 RUSSIA
\\
\vspace*{0.5cm}
E-mail: guts@univer.omsk.su  \\
\vspace*{0.5cm}
January 24, 2000\\
\vspace{.5in}
ABSTRACT
\end{center}

The topos theory is a theory which is used for deciding
a number of problems of theory of relativity, gravitation
and quantum physics.
In the article spherically symmetric
solution of  the vacuum Einstein  equations in the
Intuitionistic  theory of Gravitation at different stages
of smooth topos ${\bf Set}^{\bf L_{op}}$  is considered.
Infinitesimal "weak" gravitational  field can be strong at
some stagies, for which we have the additional dimensions.
For example, the cosmological constant is not constant
with respect to additional dimensions. Signature of space-time metric
can depend of density of vacuum and  cosmological constant.

\newpage

\setcounter{page}{1}

%%%%%%%%%%%%%%%%%%%%%%%%%%%%%%%%%%%%%%%%%%%%%%%%%%%%%%%%%%%%%%%%%%%%

This article applies the Topos theory \cite{Goldblatt} to 
theory of space-time. Other applications can be found in
\cite{Guc,Guc1,Tri,Grink,GG,Ish,Ish1}.

\section{Intuitionistic theory of gravitation}

Intuitionistic theory of gravitation is based on Synthetic
Differential Geometry of Kock-Lawvere (SDG) \cite{Kock}. SDG
is built on the base of change the field of real numbers ${\R}$ on
commutative ring ${\bf R}$, allowing to define on him differentiation,
integrating and "natural numbers". It is assumed that there exists
$D$ such that $D =\{ x\in {\bf R}~|~x^2=0 \}$ and that following
the Kock-Lawvere axiom is held:

\begin{quote}
{\it for any $g:D\rightarrow {\bf R}$ it exist
the only $a, b\in{\bf R}$ such that  $g(d)=a+d\cdot b $
for any $ d\in~D$.}
\end{quote}

This means that any function in given geometry is differentiable, but
"the law of excluded middle" is false. In other words,
intuitionistic logic acts  in SDG. But on this way one is possible
building an intuitionistic theory of gravitation in analogy with the
General theory of Relativity of Einstein \cite{Guc1,Grink,GG}.
The  elements of $d\in D$ are called infinitesimals, i.e.
infinitesimal numbers.
On the ring ${\bf R}$ we can look as on the field of real numbers $\R$
complemented by infinitesimals.

The  vacuum Einstein  equations  in SDG in space-time ${\bf R}^4$
can be written with {\it nonzero}
tensor of the energy. For instance,
$$
R_{ik}-\frac{1}{2}g_{ik}(R-2\Lambda) =
\frac{8\pi G}{c^2} \ d u_iu_k, \eqno(1)
$$
where density of matter $d\in D$ is arbitrarily taken infinitesimal
\cite{Guts1}. For infinitesimals are holding relations which are impossible
from standpoints of classical logic: $d\neq 0$,\& $d\leq ~0$ and formulas
$d=0$, $d\neq 0$ are not valied.
Such non-classical density of vacuum matter will consistent with
zero in  right part of the Einstein's equations
in the case of the vacuum in classical General theory of Relativity.
For this one is sufficiently to consider SDG in so named
well-adapted models, in which we can act within the framework of
classical logic. For instance, in smooth topos
${\bf Set}^{\bf L_{op}}$,
where ${\bf L}$ is category of {\it loci}, i.e. is the opposite
category of category of finitely generated $C^\infty$-rings \cite{Moerdijk},
the equations (1) at stage of locus $\ell A=\ell(C^\infty (\R^n)/I)$,
$I$ is a certain ideal of  $C^{\infty}$-smooth functions
from $\R^n$ to $\R$,
have the form
$$
R_{ik}(a)-\frac{1}{2}g_{ik}(a)(R(u)-2\Lambda(a))
= \frac{8\pi G}{c^2} \ d(a) u_i(a)u_k(a)\  \mod\ I, \eqno(2)
$$
where $a\in \R^n$ in parenthesises shows that we have
functions, but at stage ${\bf 1}=\ell(C^\infty(\R)/\{a\})$,
equations (2) take a classical form with null
(on $ \mod\ \{a\}$) tensor of the energy.

\section{Spherically symmetrical vacuum field}

We have the Einstein equations describing the gravitational field
created by certain material system
$$
R_{ik}-\dt\frac{1}{2}g_{ik}(R-2\Lambda)=\kappa c^2\rho u_iu_k, \eqno(3)
$$
Here $\kappa=8\pi G/c^4$, $\rho$ is density of dust in the space
which will consider further constant value.  Suppose  that dust
is described in coordinate system in which
$u_i=(\e^{-\frac{\nu}{2}},0,0,0),$
$u^k=g^{ik}u_i=(\e^{\frac{\nu}{2}},0,0,0).$
Consider case, when gravitational field possesses a central symmetry.
Central symmetry of field means that interval
of space-time can be taken in the form
$$
ds^2=\e^{\nu (r,t)}dt^2-
e^{\lambda (r,t)}dr^2-r^2(d\theta^2 +\sin^2\theta\cdot d\varphi^2)
$$
In our paper \cite{GZ} we found the following vacuum
solution of the equations (3) for which
\begin{equation}{
\rho\nu^{\prime} =0.
}
\end{equation}    %10.
1) Well-known  the   classical Schwarzschild solution when $\rho =0$
and $\nu^{\prime}\not=0$
$$
ds^2= \left(1- \dt\frac{\Lambda  r^2}{3} +
 \dt\frac{C}{r}\right)dt^2
-\frac{dr^2}
{1 - \dt\frac{\Lambda r^2}{3} + \dt\frac{C}{r} }-
$$
$$
- r^2(d\theta^2 + \sin^2\theta\cdot d\varphi^2 ). \eqno(4)
$$

2) Non-classical case when both values $\rho$ and
$\nu^{\prime}$ are simultaneously inseparable from the zero, i.e.
are infinitesimals
$$
ds^2= \left(1+ \dt\frac{(\kappa c^2\rho -2\Lambda ) r^2}{6} +
 \dt\frac{C}{r}\right)dt^2
-\frac{dr^2}
{1 - \dt\frac{(\Lambda + \kappa c^2\rho )r^2}{3} + \dt\frac{C}{r} }-
$$
$$
- r^2(d\theta^2 + \sin^2\theta\cdot d\varphi^2 ) \eqno(5)
$$
This metric can be called the  non-classical Schwarzschild  solution
of the Einstein equations. But here $\Lambda$ and $C$ are infinitesimals.
So (5) is {\it infinitesimal} gravitational field, which is not weak
in some sence (see below \S 5).

3) Suppose  that gravitational field has no singularity in all space.
This means that metric has no singularity in $r=0$. So we shall
consider that $C$ is  zero. But it is proved that
$$
2\Lambda\rho = \kappa c^2\rho^2 \eqno(6)
$$
 and, besides, $\Lambda$ is inconvertible
value of ring ${\bf R}.$

% Thence follows that density of matter $\rho$ is inconvertible value.

In other words, matter has non-classical density,
and its gravitational field has the form
$$
ds^2= \left(1+ \dt\frac{(\kappa c^2\rho -2\Lambda ) r^2}{6}
\right)dt^2
 -\frac{dr^2}
{1 - \dt\frac{(\Lambda + \kappa c^2\rho )r^2}{3} }
- r^2(d\theta^2 + \sin^2\theta\cdot d\varphi^2 )  \eqno(7)
$$
Below we consider solutions (5), (7) in models which are
different toposes, or, more exactly in  categories of presheaves
${\bf Set}^{{\cal C}^{op}},$ where ${\cal C}$ is such
subcategory of category ${\bf L}$ \cite{Moerdijk} that
ring ${\bf R}={\rm Hom}_{{\bf L}}(-,\ell C^\infty (\R))$
is local and Archimedean
\footnote{for example, categories of closed and germ-detrmined ideals are
such subcategories \cite{Moerdijk}}.

\section{Stagies}

Synthetic Differential Geometry uses "naive" style, i.e. contains
term "element", or the set-theoretic formulas of the form $a\in A$.
So it is necessary
to be able to understand this naive
writing as referering to cartesian closed categories, and to toposes in
particular, because the Kock-Lawvere axiom has no models in the
category of sets. The method for deciding this problem is introducing
of {\it generalized element $b\in_XB$}, that is, a map $X\to B$,
where $X$ is an arbitrary object, called the {\it stages of definition},
or the {\it domain of variation} of the element $b$. The "classical" element
is a map ${\bf 1}\to B$. As A.Kock writes "when thinking in terms of
physics (of which geometry of space forms a special case), reason for
the name "domain of variation" (instead of "stage of definition") becomes
clear: for a non-atomistic point of view, a body $B$ is not described
just in terms of its "atoms" $b\in B$, that is ${\bf 1}\to B$, but
in terms of "particles" of varying size $X$, or in terms of motions that
take place  in $B$ and are parametrized by a temporal extent $X$; both
of these situations being described by maps $X\to B$ for suitable
domain of variation $X$ \cite{Kock}. In our case
the role of "body" $B$ will play a
gravitational vacuum field $g_{ik}$, or geometry of space-time,
which we will study at different stages, or
with respect to different points of view
concerning the possible geometric structure of the World.

In the case of topos  ${\bf Set}^{\bf L}$ the concept of stage is realized
with the help of the folowing method.

There exists the Yoneda embeding \cite[p.26]{MacLane}
$$
y: {\bf L} \to {\bf Set}^{{\bf L}^{op}}, \ \
$$
$$
y(\ell A)(\ell B)={\rm Hom}_{{\bf L}}(\ell B,\ell A)
$$
and for a morphism $\alpha:\ell B\to\ell C$
$$
y(\ell A)(\alpha): {\rm Hom}_{{\bf L}}(\ell C,\ell A) \to
{\rm Hom}_{{\bf L}}(\ell B,\ell A),
$$
$$
y(\ell A)(\alpha)(u)=u\circ \alpha, u:\ell C \to\ell A.
$$
Briefly for the Yoneda embedings we write
$$
y(\ell A)={\rm Hom}_{{\bf L}}(-,\ell A).
$$
Instead of $y(\ell A)$ we shall write simply $\ell A$.
So if ring ${\bf R}$ is $\ell C^\infty (\R)$ then
$$
{\bf R}\equiv y({\bf R})= {\rm Hom}_{{\bf L}}(-,\ell C^\infty (\R) ).
$$
Hence element of ring ${\bf R}$, i.e. intuitionistic {\it real number} can
be represented by arbitrary morphism of the form
$\ell A\to \ell C^\infty (\R)$. We say in such case that one have
{\it real at stage} $\ell A$. It means that metric (5) must be considered
at different stages. For example, at stage $\ell A=\ell C^\infty (\R^n)/I$,
where $I$ is some ideal of ring $C^\infty (\R^n)$.

Note that an event $x $ of the space-time ${\bf R}^4$ at stage
        $\ell A$  is the class of
        $C^{\infty}$-smooth vector functions
        $(X^0(a),X^1(a),X^2(a),X^3(a)):\R^n\rightarrow  \R^4$,
where each function     $X^i(a)$ is taken by $\mod \ I$.
     The argument $a\in\R^n$ is some "hidden" parameter corresponding
        to the stage $\ell A$. Hence it follows that at stage of real
        numbers ${\bf R}=\ell C^{\infty}(\R)$ of the topos
under consideration  an event $x$ is described by just a
        $C^{\infty}$-smooth vector function
        $(X^0(a),X^1(a),X^2(a),X^3(a)), a\in \R$.
	At stage of ${\bf R}^2=\ell C^{\infty}(\R^2)$
        an event $x$ is 2-dimensional surface, i.e. a {\it string}.
      The classical four numbers $(x^0,\ x^1,\ x^2,\ x^3)$, the
coordinates of the event $x$,
        are obtained at the stage
${\bf 1}=\ell C^{\infty}(\R^0)= \ell C^{\infty}(\R)/\{a\}$
    (the ideal $\{a\}$ allows one to identify functions if their values at
        $0$ coincide), i.e., $x^i=X^i(0), i=0,1,2,3$.

There exists a number of types of infinitesimals:  first-order
infinitesimals
$D=\{x\in {\bf R}| x^2=0\}$, $k^{th}$-order infinitesimals
$D_k =\{x\in {\bf R}| x^{k+1}=0\}$, and the infinitesimals
$\d=\{x\in {\bf R}| f(x)=0, all f\in m^g_{\{0\}}\}$, where
$m^g_{\{0\}}$ is the ideal of functions having zero germ at $0$,
i.e. vanishing in a neighborhood of $0$,
$$
D\subset D_2\subset\dots\subset D_k\subset\dots\subset \d.
$$

Infinitesimals $\rho, \Lambda\in\d$ at stage
$\ell A=\ell C^\infty (\R^n)/I$
are the classes $\rho (a)\ \mod\ I$, $\Lambda (a)\ \mod\ I$
such that for every
$\phi\in m^g_{\{0\}}$, \ $\phi\circ\rho\in I$, \ $\phi\circ\Lambda\in I$,
where $\rho:\R^n\to\R$ \cite[p.77]{Moerdijk}, or
$$
(\phi\circ\rho)(a)=0\ \mod\ I, \qquad (\phi\circ\Lambda)(a)=0\ \mod\ I
$$
The condition (6) has the form
$$
2\Lambda(a)\rho(a)-\kappa c^2\rho^2(a)=0\ \mod\ I.   \eqno(6')
$$

Let us to consider the forms of metrics (5), (7) at different stages.
%Each stage is some variant of theory of gravitation.

\subsection{Stage ${\bf 1}$}

For classical General
Theory of Relativity we have stage
${\bf 1}=\ell C^\infty (\R^0)=\ell C^\infty (\R)/\{a\}$. At this
stage metric
(7) is metric of Minkowski space-time
$$
g_{ik}(a)=\left(
\begin{array}{cccc}
 1 &0 &0 &0\\
0 &-1 &0 &0\\
0 &0 &-r^2\sin^2\theta &0\\
0 &0 &0 &-r^2
\end{array}
\right),
$$
i.e. $\Lambda$ and $\rho$ are equal to zero.
In fact,
$$
\phi(\rho (a))=\phi(\rho (0))+ \phi^{\prime}(\rho (0))
\rho^{\prime} (0)a+o(|a|).\eqno(8)
$$
Since $\phi(\rho (a))\in I=\{a\}$ then $\phi(\rho (0))=0$.
Hence $\rho (0)=0$, because $\phi\in m^g_{\{0\}}$.
Then $\rho \ \mod\ I=0$. Simiraly $\Lambda \ \mod\ I=0$,
$C \ \mod\ I=0$

\subsection{Stage {\bf $D=\ell C^\infty(\R)/\{a^2\}$ }}

In this case
$$
g_{ik}(a)=
$$
$$
\left(
\begin{array}{cccc}
 1+\dt\frac{(\kappa c^2\rho_1 -2\Lambda_1)}{6}
 \cdot a r^2 +\frac{C_1a}{r}&0 &0 &0\\
0 &-\dt\frac{1}{1-\dt\frac{(\kappa c^2\rho_1+\Lambda_1)}{3}
\cdot a r^2 +\frac{C_1a}{r}} &0 &0\\
0 &0 &-r^2\sin^2\theta &0\\
0 &0 &0 &-r^2
\end{array}
\right)
$$
In fact, it follows from (8) that
$\phi(\rho (0))=\phi^{\prime}(\rho (0))\rho^{\prime} (0)=0$. Since
$\phi |_U\equiv 0, \ \phi^{\prime} |_U\equiv 0$ for some
neighborhood of $0$,
then $\rho (0)=0$. So $\rho \ \mod I=\rho_1 a$, $\rho_1 \in\R$.
Similaly  $\Lambda\ \mod\ I =\Lambda_1a$, $\Lambda_1\in \R$.

\subsection{Stage {\bf $D_p=\ell C^\infty(\R)/\{a^{p+1}\} $ }}

Here
$$
g_{00}(a)= 1+\sum\limits_{k=1}^{p}\left[\dt\frac{
(\kappa c^2\rho_k -2\Lambda_k)}{6}
    \cdot  r^2+\frac{C_k}{r}\right]a^k
$$
$$
g_{11}(a)= -\frac{1}{ 1-\sum\limits_{k=1}^{p}\left[
\dt
\frac{ (\kappa c^2\rho_k+\Lambda_k) }{3}\cdot  
r^2- \frac{C_k}{r} \right] a^k}
$$
and others $g_{ik}$ are classical.

\subsection{Stage {\bf $D_n(k)=\ell J_n^k=\ell C_0^\infty(\R^n)/m^{k+1}$ }}

Let $m={f|f(0)=0}$ is maximal ideal of ring $\ell C_0^\infty(\R^n).$
Then
$$
g_{00}(a,t,r,\varphi,\theta)= 1+\sum\limits_{l=1}^{k}
\sum\limits_{i_1,\ldots,i_l=1}^{n}\left[
   \dt\frac{(\kappa c^2\rho_{i_1,\ldots,i_l} -2\Lambda_{i_1,\ldots,i_l})}{6}
    r^2  +\frac{C_{i_1,\ldots,i_l}}{r}  \right] a_{i_1}\ldots a_{i_l}   \\
$$
$$
g_{11}(a,t,r,\varphi,\theta)= -\dt\frac{1}{1-\sum
\limits_{l=1}^{k}\sum\limits_{i_1,\ldots,i_l=1}^{n}\left[
      \dt\frac{(\kappa c^2\rho_{i_1,\ldots,i_l}+
\Lambda_{i_1,\ldots,i_l})}{3}
   r^2 - \frac{C_{i_1,\ldots,i_l}}{r} \right]a_{i_1}\ldots a_{i_l}  } \\
$$
Here $a=(a_1,\ldots,a_n)$.

\subsection{Stage {\bf $\ell C^\infty(\R^2)/\{a_1-a_2\}$ }}

In this case functions $\Lambda$, $\rho, C$ depend of one variable,
for example, $a_2,$ and vanishing at $0$. Then
$$
g_{00}=1+\dt\frac{ \kappa c^2\rho(a_2) -2\Lambda(a_2) }{6} r^2 +
\frac{C(a_2)}{r}
$$
$$
g_{11}= -\dt\frac{1}{ 1-\dt\frac{ \kappa c^2\rho(a_2)+\Lambda(a_2) }{3} r^2
+\frac{C(a_2)}{r} }
$$

\subsection{Stage {\bf $\ell C^\infty(\R)/\{\sin\pi a, \cos\pi a\}$ }}

If $\rho\in D$ then
$$
\rho(a)=\frac{\alpha_0}{2}+\sum\limits_{k=0}^{\infty}\alpha_k\cos k\pi a
+\beta_k\sin k\pi a = A \ \mod \ I, \qquad A\in\R,
$$
$$
\rho^2(a)=A^2\ \mod \ I \in I \Longrightarrow A=0.
$$
Hence $\rho=0$, $\Lambda=0$ and $g_{ik}$ in this case coincides with
Minkowski metric.

\subsection{Stage {\bf $\ell C^\infty(U)$ }  }

Consider stage  $\ell C^\infty(U)$, where $U\subset \R^n$ is open set.
Since
$$
\ell C^\infty(U) \cong  \ell C^\infty(\R^{n+1})/\{a_{n+1}\cdot \chi(a)-1\},
$$
$$
U=\{a\in \R^n|\chi(a)\neq 0\}, \qquad \chi \in C^\infty(\R^n),
$$
then with the help of transformation of variables
$$
\left\{
\begin{array}{lcl}
a\prime=a\\
a_{n+1}^\prime=a_{n+1}\cdot\chi(a)-1\\
\end{array}
\right.
$$
we can get, for example, that
$$
\rho (a,a_{n+1})\ \mod I = \sum\limits_{|\alpha|=1}^{\infty}A_{\alpha}
a^{\alpha},
\qquad a\in U, \quad A_\alpha\in \R
$$
$$
\rho (0,\frac{1}{a(0)})=0.
$$

\section{Transitions between stages}

Change of stage $\ell A$ on stage $\ell B$ is morphism between two stages
$$
\ell B\stackrel{\psi}{\rightarrow}\ell A
$$
then transition between $Hom_{{\bf L}}(\ell A,T)$ $Hom_{{\bf L}}(\ell B,T)$
is realized by means of
$$
Hom_{{\bf L}}(\ell A,T)\stackrel{\Psi}{\rightarrow}
Hom_{{\bf L}}(\ell B,T)
$$
for any object $T$ of category ${\bf L}$,
which each morphism $h:\ell A\rightarrow T$ puts in correspondence
morphism $h\Psi:\ell B\rightarrow T$.

Let, now, $\ell A=\ell C^\infty(\R^n)/I$ and
$\ell B=\ell C^\infty(\R^m)/J.$ Then transition between stages
$\ell A$, $\ell B$ gives metric
$$
g_{ik}(b)=
$$
$$
 \left(
\begin{array}{cccc}
 1+\dt\frac{(\kappa c^2\rho(\Psi(b)) -2\Lambda(\Psi(b)))}{6} r^2 &0 &0 &0
\\ 0 &-\dt\frac{1}{1-\dt\frac{(\kappa c^2\rho(\Psi(b))
+\Lambda(\Psi(b)))}{3} r^2 } &0 &0\\
0 &0 &-r^2\sin^2\theta &0\\

0 &0 &0 &-r^2
\end{array}
\right)
$$
modulo $J\cdot C^\infty(\R^m\times\R^4)$ instead of metric (7).

The condition of infinitesimallity for $\Lambda$, $\rho$ and
condition (6) will copied so
$$
(\phi\circ\rho\circ\Psi)(b)=0\ \mod J, \qquad
(\phi\circ\Lambda\circ\Psi)(b))=0\ \mod J
$$
and
$$
2\Lambda(\Psi)(b))\rho(\Psi)(b))-\kappa c^2\rho^2(\Psi)(b))=0\ \mod\ J.
$$

\section{Physical notes}

At first note very interesting fact: at all considered stages
signature of metric
$g_{ik}$ depends of the form of functions $\Lambda$,  $\rho$  and 
$C$. For example,
at stage $D=\ell C^\infty(\R)/\{a^2\}$ \ \ $\rho \ \mod I=\rho_1 a$,
$\Lambda\ \mod\ I =\Lambda_1a$, $C\ \mod\ I =C_1a$, where
$\rho_1, \Lambda_1, C_1\in \R$ are
arbitrary real numbers (under $C=0$ the condition (6) or
(6') is valid for all $a\in\R$). Hence field $g_{ik}$ is not weak with
respect to five dimensions $(t,r,\theta,\phi,a)$. More interesting situation
can be observed at stages $D_n(k)=\ell J_n^k=lC_0^\infty(\R^n)/m^{k+1}$
and $\ell C^\infty(\R^2)/\{a_1-a_2\}$.

Note also that at stage $\ell C^\infty(U)$ if $U$ is bounded then functions
$\rho, \Lambda, C$ are can be taken small and signature of metric
does not change.

What sence have the "hidden" parameters $a\in\R^n$? We are thinking
that they say us about existence of the additional dimensions the number
of which can be changeable.
For finding of coefficients $\rho_1, \Lambda_1, C_1$ at stage
$\ell C^\infty(\R)/\{a^2\}$,
$\rho_{i_1,\ldots,i_l},\Lambda_{i_1,\ldots,i_l}, C_{i_1,\ldots,i_l}$
at stage $\ell J_n^k=lC_0^\infty(\R^n)/m^{k+1}$, functions
$\rho(a_2), \Lambda(a_2), C_(a_2)$ at stage
$\ell C^\infty(\R^2)/\{a_1-a_2\}$ must be possibly used the
many-dimensional Einstein equations. In other words
4-dimensional intuitionistic theory contains uncountable number
of many-dimensional theories. The infinitesimal field with respect to
4-dimensional universe can be found non-weak with respect to
"hidden" geometry.  Intuitionistic logic implies new view about
Nuture of World.


{\small
\begin{thebibliography}{99}

\bibitem{Goldblatt}
Goldblatt, R. Topoi. The categorical analisys of logic.
Amsterdam: North-Holland Publ. Comp., 1979.

\bibitem{Kock}
Kock A. Synthetic Differential Geometry. Cambridge University Press,
1981.

\bibitem{Guc}
Guts, A.K. A topos-theoretic  approach  to
the  foundations  of Relativity theory // Soviet Math.
Dokl.- 1991.-V.43, No.3.-P.904-907.
\bibitem{Tri}
Trifonov, V. Linear Solution of the Four-Dimensionality
Problem // Europhys. Lett. 1995. V.32, N.8, P.621-626.

\bibitem{Guc1}
Guts, A.K. Intuitionistic Theory of space-time //
International geometric school-seminar of memory N.V.Efimov. Avstracts.
-- Rostov-on Don, 1996. P.87-88.

\bibitem{Grink}
Grinkevich, E.B.. Synthetic Differential Geometry: A Way to Intuitionistic
Models of General Relativity in Toposes -- Paper gr-qc/9608013

\bibitem{GG}
Guts, A.K., Grinkevich E.B. Toposes in General Theory of Relativity. --
Paper gr-qc/9610073.


\bibitem{Ish}
Isham, C.J. Topos Theory and Consistent Histories:
The Internal Logic of the Set of all. -- Paper gr-qc/9607069

\bibitem{Ish1}
Isham, C.J. Some possible Roles for Topos Theory in Quantum Theory
and Quantum Gravity. Paper gr-qc/9910005. 

\bibitem{Guts1}
Guts, A.K. Vacuum equations of Einstein in Synthetic Differential Geometry
of Kock-Lawvere. Abstracts of X Russian gravitational conferences,
Vladimir, 1999.

\bibitem{Moerdijk}
Moerdijk, I., Reyes, G.E. Models for Smooth Infinitesimal Analysis.
Springer-Verlag, 1991.

\bibitem{GZ}
Guts, A.K., Zvyagintsev, A.A. Solution of  the vacuum Einstein  equations
in Synthetic Differential Geometry of Kock-Lawvere.
-- Paper physics/9909016.

\bibitem{MacLane}
MacLane, S., Moerdijk, I. Sheaves in Geometry and Logic.
Springer-Verlag, 1994.

\end{thebibliography}
}
\end{document}